\documentclass[preprint]{revtex4-1} 

\usepackage{amsfonts} 
\usepackage{graphicx}
\usepackage{amsmath, amssymb}
\usepackage{slashed}
\usepackage{braket}
\usepackage[bottom]{footmisc}
\graphicspath{  {images/} }
\begin{document}


\title{Implications of a holographic density of states on inflation}

\author{Satish Ramakrishna}
\email{ramakrishna@physics.rutgers.edu}
\affiliation{Department of Physics \& Astronomy, Rutgers, The State University of New Jersey, 136 Frelinghuysen Road
Piscataway, NJ 08854-8019}


\date{31-Jan-2022}

\begin{abstract}

There is theoretical evidence that the number of degrees of freedom in quantum fields decreases as one studies them at extremely short distances. This emerges from the study of entropy of black holes, as well as from holographic theories in AdS geometries.  Presumably a theory of quantum gravity will provide an explicit description of how the number of degrees of freedom thin out as one studies high energy scales.

We do not have a comprehensive theory of how such a thinning of degrees of freedom would occur. It is likely that there might be residual (and measurable) effects at larger length scales, though this might be significant only near the Planck scale. There are very few instances in Nature where one might be able to see effects of this thinning. One promising venue is in the phenomenon of inflation, produced in the simplest models through a scalar inflaton field in a potential with a flat (``slow-roll'') part as well as a potential well. During inflation, fluctuations of small length scales are stretched to large scales and then exit the Hubble horizon. We compute the effect of such a thinning of degrees of freedom upon the running of the spectral index of quantum fluctuations of the inflaton and deduce that this will lead to a small positive power of wave-vector (opposite to the usual $\sim -\epsilon$, i.e., negative power correction). Some comments are then made about the impact on observations (or non-observations) of such fluctuations \cite{Martin}.

\end{abstract}
\maketitle 

\section{Introduction}

The AdS/CFT correspondence \cite{Dawid, Susskind, Maldacena} as well as various hints we have from the study of entropy of quantum systems \cite{Bekenstein, Srednicki} appears to drive towards a conclusion that space-time is holographic. A related idea, that of thinning of the density of states available for fluctuations in a quantum field, has been explored, for instance by Banks \cite{Banks1} - though this idea has been explored in various parts of the literature in different contexts \cite{Weinberg, Reuter1, Percacci, Reuter2}. In concrete terms, this would mean that the number of degrees of freedom in a quantum field, for instance, would decrease, i.e., thin out, from the expected $(3+1)$-dimensional to lower-dimensional at short distances.  This thinning would be expected to be continuous and should be a result of a quantum theory of space-time and gravity. It would be manifested as the measure in Fourier space modes $d^3k \: d\omega$ changed to $d^{3-q}k \: d \omega$ when computing loop diagrams in quantum-field-theory. Specifically, one would expect $q$ itself to be a function of $k$ - an example of such a behavior might be as in Fig. 1, where $G$ would be a scale where the density of states drops by one power of momentum above the scale $G$ - we have plotted the function $q=\frac{4-D}{2}$ where $D$ is the number of space-time dimensions. However, one would expect $q$ to differ from $0$ in some range around $G$, which might conceivably extend a few orders of magnitude below the scale $G$. An example is shown in Fig. 1 in a Log-Linear Plot.

\begin{figure}[h!]
\caption{Approximation for q(k/G)}
\centering
\includegraphics[scale=.5]{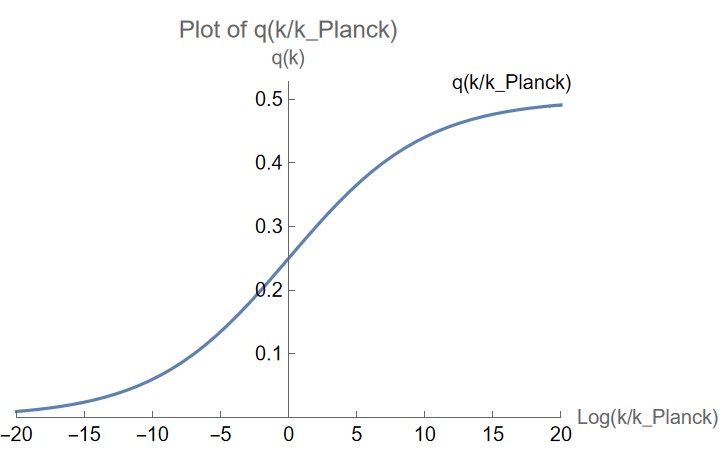}
\end{figure}

We see this method as an approximation technique, much along the lines of dimensional regularization. The key consequence of semi-classical quantum gravity, for instance the Bekenstein \cite{Bekenstein} formula is that there are fewer degrees of freedom in a given patch of spacetime than would be estimated in quantum field theory. That can be modelled by a modification of the measure used to perform integrals in quantum field theory. The theory of inflation naturally involves very short-distance physics – we make no claim to constructing a quantum theory of gravity, merely a simulation of one, by modeling the reduction in degrees of freedom of the quantum field.


A key point is that, like dimensional regularization, we apply this approximation method in momentum-space integrals. The corresponding logic is not intuitive at all in real-space. Hence, we write the action (see Equation (1)) in real-space, then re-write it in momentum-space, followed by applying the condition that the degrees of freedom are a function of momentum.


The various theories of inflation \cite{Riotto} usually incorporate scalar fields (one or many) that have quantum fluctuations at all scales during inflation, while their corresponding length scales are under the then Hubble horizon. These fluctuations are then frozen, as these modes exit the Hubble horizon (see Fig. 2) and they are then seen when they re-enter the horizon billions of years later. These quantum fluctuations occur at all length scales and are expanded by a factor $\frac{a(t_0)}{a(t_I)}$, where $a(t_0)$ is the current scale factor, while $a(t_I)$ is the scale factor at the start of inflation. This number is $\sim 10^{40}$ at least (the uncertainty is in the exponential) and hence very short wavelength fluctuations (Planck scale) are massively expanded in size (by this factor), then, they re-enter the horizon. These fluctuations can be either scalar or tensor in nature and a rather simple prediction from the simplest theories of inflation is that the scalar modes (as a function of momentum) are very close to scale invariant. A spectral index, defined from $P_{\chi} \sim k^{n_s-1}$ (the power spectrum \cite{Riotto}) is used to characterize the deviation from a scale-invariant ($n_s=1$) prediction. From experimental observations\cite{Wang}, $n_s-1 \approx -0.04$.

We ask the natural question - if the density of states of a scalar field (in the simplest theories of inflation) is indeed constrained for high-momentum/short-wavelength modes, can we see this in the fluctuation in a contribution to the spectral index? If so, what is the form of this contribution and does the result exclude some modifications to the density of states.

The question is complicated by the fact that we have a variety of theories of inflation and so a number of possible contributions to the spectral index. In that case, {\it {it is possible that the density of states formulation we choose, constrains (and is constrained by), the kinds of theories of inflation that would be consistent with it}}.

\begin{figure}[h!]
\caption{Momentum modes entering and exiting the Hubble horizon}
\centering
\includegraphics[scale=.5]{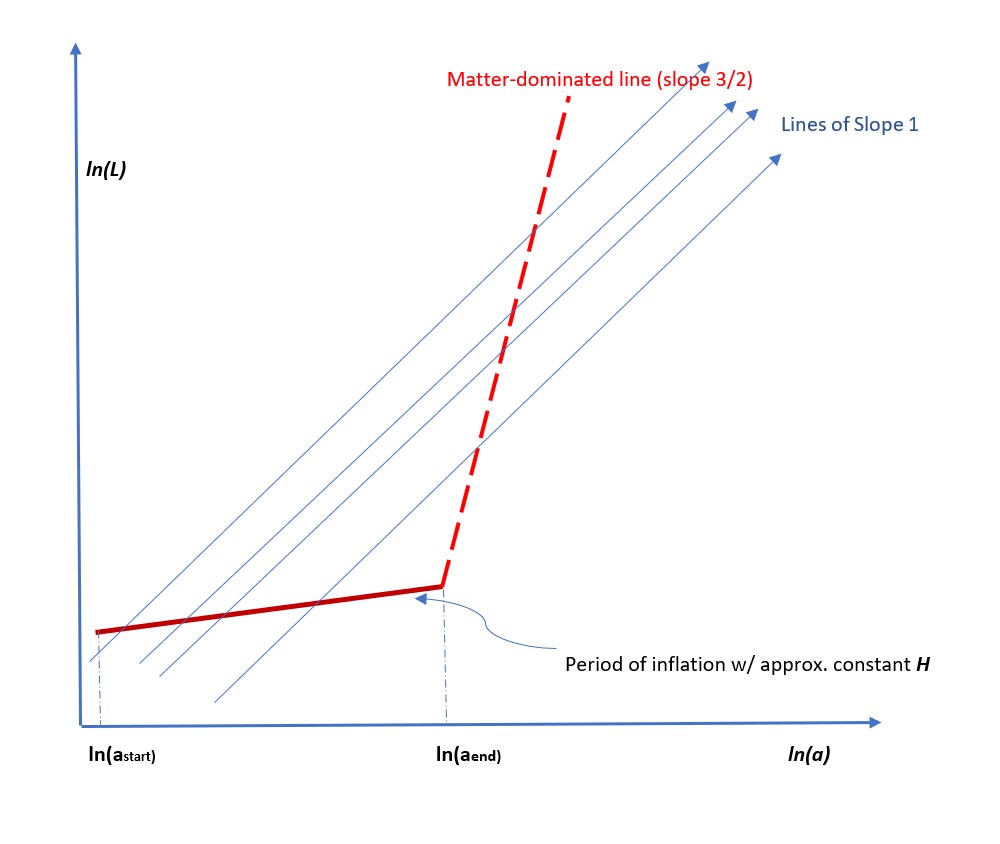}
\end{figure}

\subsection{The plan of the calculation}

We carry out the calculation for inflation caused by a single scalar field. We write down the action and then compute the equation obeyed by the Fourier modes in $D$-dimensions. We then proceed to solve this equation for modes whose wavelengths are much smaller than and also much larger than the Hubble horizon. The full solution is then obtained by matching the two limiting cases - this allows us to write down the full solution. We then use this to compute the power spectrum for fluctuations for the various modes and thus the spectral index, i.e., the $k$-dependence of the power spectrum. 

This calculation is a prelude to the full computation of the power spectrum of scalar perturbations to the metric from such a scalar field. This calculation is performed in a convenient gauge (the flat gauge) and the solution is used to compute the power spectrum of the fluctuations. We thus obtain the spectral index for these scalar modes. The scalar modes are directly observed in the angular dependence of the cosmic-microwave-background (CMB) radiation and the spectral index can be calculated from the observed data.  This spectral index is compared to what we would expect.

In each case, we let the parameter $q=\frac{4-D}{2}$ be slightly different from $0$ and judge the impact upon the spectral index. This can be directly compared to observation and useful conclusions may be drawn about excluding some proposals for the density of high momentum states in quantum-field-theory. 

\subsection{Comparison to previous analyses}

Previous work \cite{Brandenberger1, Danielsson, Cheung} has been done on similar aspects in this area (study of the spectral index). In particular, Martin \& Brandenberger1 (hereafter referred to as M\&B) study non-linear dispersion relations, i.e., where the action studied in the next section is quite non-linear for large momentum. In such a situation, due to this large-$k$ behavior, the ground state does not admit of plane wave solutions - in particular,  (M\&B)  refer to three regions of the solutions, {\it viz.,} 
\begin{enumerate}
\item Region I: $k >> \frac{1}{l_{Pl}}$ where $l_{Pl}$ is the Planck length
\item Region II: $ \frac{1}{l_{Pl}} > k >H$ where $H$ is the Hubble constant (inverse Hubble radius)
\item Region III: $k << H$ which is well past the point where the modes exit the Hubble horizon
\end{enumerate}
As it turns out, all relevant parameters and scaling behavior (i.e., the spectral index) are found from matching between regions. In Region III, we find, in all generality, with the modification to the measure discussed in the above, corrections to the spectral index proportional to $q$ (with $q(k)$ measured at the relevant scale, see Section III). This is purely a consequence of the matching condition and the nature of the solution to the problem, so it appears as a correction to all the cases that are considered in the references mentioned above, even though the ground states for large-$k$ are very different in all the non-linear cases. Concretely, in all the cases in  (M\&B), the spectral index will have the $q$-related correction discussed below.

The studies of Danielsson \cite{Danielsson} concern corrections to Equation (10) in the following section due to adiabatic evolution from a standard vacuum at some particular conformal time. While there is no non-linearity here, the same matching conditions lead to corrections exactly as below in the spectral index.

\section{Calculation of spectral index for a scalar field in $D$-dimensions}

We start with the action for a scalar field in $D$-dimensional space-time ($D-1$-space dimensions and $1$ time dimension). We work with the metric signature $(+---)$ and  set $\hbar=c=1$. We work with an FLRW-metric in conformal time (written as $\tau$). The dimensionless scale factor is the function $a(\tau)$. The Hubble parameter $H$ is, as usual $\frac{\dot a}{a}$ where $\dot a$ is the {\it {time}} derivative of the scale factor. The action integral is
\begin{eqnarray}
{\cal S}[\chi] = \frac{1}{2} \int d^{D-1}x \: d \tau \: \sqrt{- det \: g} \: \: g^{\mu \nu} \partial_{\mu} \chi \partial_{\nu} \chi
\end{eqnarray}
Here
\begin{eqnarray}
det \: g = a^{2D} \: \: \: , \: \: \:  g^{\mu \nu} \partial_{\mu} \chi \: \partial_{\nu} \chi = \frac{1}{a^2} \bigg( (\chi^{'})^2 - (\nabla \chi)^2 \bigg) \: \: \: , \: \: \:  f^{'} \equiv \frac{\partial f}{\partial \tau}
\end{eqnarray}
Next, we define $q = \frac{4-D}{2}$ and we set $\chi = \frac{\sigma}{a^{\frac{D-2}{2}}}=\frac{\sigma}{a^{1-q}}$ and rewrite the action as
\begin{eqnarray}
{\cal S}[\sigma] = \frac{1}{2} \int d^{D-1}x \: d \tau \: \bigg(  (\sigma^{'})^2+\sigma^2 \bigg(\frac{a^{'}}{a} \bigg)^2 \bigg(\frac{D-2}{2}\bigg)^2 -  (\nabla \sigma)^2 - \frac{D-2}{2} (\sigma^2)^{'} \bigg(\frac{a^{'}}{a} \bigg) \bigg)
\end{eqnarray}
Integrating by parts allows us to write this action as
\begin{eqnarray}
{\cal S}[\sigma] = \frac{1}{2} \int d^{D-1}x \: d \tau \: \bigg[  (\sigma^{'})^2+\sigma^2 \bigg(\frac{a^{'}}{a} \bigg)^2 \bigg(\frac{D-2}{2}\bigg) \bigg(\frac{D-4}{2}\bigg)-  (\nabla \sigma)^2 + \frac{D-2}{2} \sigma^2 \frac{a^{''}}{a} \bigg]
\end{eqnarray}

Now, we write the action integral in momentum space and then set $q = \frac{4-D}{2}$ to be a function of momentum $k$. We can then write down the equation for the Fourier modes\cite{Comment} of $\sigma$ as 
\begin{eqnarray}
\sigma_k^{''} + \bigg( k^2-(1-q(k) \: ) \frac{a^{''}}{a} + q(k) \:  (1-q(k) \: )\bigg(\frac{a^{'}}{a} \bigg)^2  \bigg) \sigma_k = 0
\end{eqnarray}

This equation can be exactly solved, though a rather more intuitive way to do so is to realize that $a(\tau) = - \frac{1}{H \tau}$ for de-Sitter evolution and $\frac{a^{''}}{a} \sim \bigg(\frac{a^{'}}{a} \bigg)^2  \sim \frac{1}{\tau^2}$, so that the regimes $|k \tau| = \frac{k}{a H}$ is either much larger or smaller than 1 are interesting, with a matching condition at $k =a H$.

In the limit $k>>a H$, which corresponds to sub-Hubble-horizon length scale fluctuations of $\sigma$, the equation (in the Bunch-Davies \cite{Riotto, Davies} vacuum with the usual normalization) can be solved to give
\begin{eqnarray}
\sigma_k(\tau) = \frac{1}{\sqrt{2 k}} e^{-i k\tau}
\end{eqnarray}

Note that, in line with the analysis in M\&B\cite{Brandenberger1}, the Unruh dispersion relation \cite{Unruh} and a different choice of vacuum, determine that the relevant $\sigma_k \sim \frac{1}{k}$ in our notation. That affects the result they get (and for this analysis too), as described below.

In the opposite limit, $k <<a H$, which corresponds to super-Horizon-length scale fluctuations, there is a growing solution to the equation
\begin{eqnarray}
\sigma_k^{''} = \bigg( (1-q) \frac{a^{''}}{a} - q (1-q)\bigg(\frac{a^{'}}{a} \bigg)^2  \bigg) \sigma_k 
\end{eqnarray}
The solution can be checked to be $\sigma_k \sim a^{1-q}$. This amounts to a correction to the usual ``freeze-out'', where the fluctuations increase in sync with expansion - note that $\chi_k = \frac{\sigma_k}{a^{1-q}}$, so the $\chi_k$ modes are indeed frozen out. As (a correction to the) usual, the solution can be written as $\sigma_k = B(k) a^{1-q}$ and we obtain $B(k)$ through matching the solutions at $k=aH$. This yields
\begin{eqnarray}
B(k) = \frac{H^{1-q}}{\sqrt{2k^3}} k^q
\end{eqnarray}

Here, $q$, which is a function of the momentum $k$, is then set to be the value at $k=aH$, so we are sensitive to the dimensional effects at the scale of the Hubble horizon.

In the other (Unruh vacuum) case studied by M\&B \cite{Brandenberger1}, $B(k) \sim k^{q-2}$.

Hence, the solution for the original $\chi_k$ mode (in Fourier space) is
\begin{eqnarray}
\chi_k=\frac{H^{1-q}}{\sqrt{2k^3}} k^q
\end{eqnarray}
This reduces to the usual solution if $q=0$.

In the reference mentioned in Section IIA \cite{Brandenberger1}, various other vacua with non-linear dispersion relations are studied. They lead to different matching conditions, however, they all are corrected by the same power of $q(k=aH)$, since that derives from the small-$k$ solution, which  is known to be in the regime of linear dispersion anyway and Equation (7) applies.

We can now compute the power spectrum of the scalar field. It would be
\begin{eqnarray}
{\cal P}_{\chi}(k) = \frac{k^3}{2 \pi^2} |\chi_k|^2 \nonumber \\
= \bigg(\frac{H^{1-q}}{2 \pi} \bigg)^2 k^{2 q}
\end{eqnarray}

And, in parallel, for the Unruh vacuum case, ${\cal P}_{\chi}(k) \sim k^{2q-1}$.

We could generalize the above discussion for a slight variation of the Hubble parameter for different modes \cite{Riotto} and obtain the spectral index
\begin{eqnarray}
n_{\chi} - 1 = 2q-2 (1-q)\epsilon
\end{eqnarray}

This is one of two central results of this paper. There is a modification to the spectral index due to this effect, for non-zero $q$. For zero $q$, the result reduces to the usual result for a massless scalar field. As described earlier, $q=q(k=aH)$.

In parallel with the above analysis, which studies the behavior with the Bunch-Davies vacuum, the Unruh vacuum result would be 
\begin{eqnarray}
n_{\chi} - 1 = 2q-1 - 2(1-q) \epsilon
\end{eqnarray}

\section{Calculation of the spectral index for scalar perturbations with non-zero $q$}

In this case, we simply follow \cite{Riotto} and study the comoving (scalar) curvature perturbation $\zeta$. The theory of perturbations of the scalar field in a background metric is well developed - the general metric we work with is explained thoroughly in various introductory papers and texts \cite{Brandenberger2, Riotto, Baumann}
\begin{eqnarray}
(ds)^2=(1+2 \Phi) d\tau^2-2 B_i d\tau d x^i - ((1-2 \psi) \delta_{ij} + (\partial_i \partial_j-\frac{1}{D-1} \delta_{ij} \partial^2E) dx^i dx^j
\end{eqnarray}
In the above, $\Phi$ is called the lapse function (or the gravitational potential), $\psi$ , $E$, $B_i$ are all as usually defined to obtain scalar metric perturbations. The metric has some gauge freedom, since physics is insensitive to general coordinate transformations (diffeomorphisms) \cite{Baumann}. Hence, one usually works in a particular gauge.

We work in the flat gauge, with zero lapse function and work to lowest order in fluctuations, so we neglect the graviational potential fluctuation. The $D$ related terms do not change the result obtained {\it vs.} the usual computation since we are considering only scalar perturbations and further, work in a gauge that drops even them (the computation for the tensor terms is substantially more involved). In this gauge, with fluctuations in the scalar field $\delta \chi$,
\begin{eqnarray}
\zeta = H \frac{\delta \chi}{\dot \chi} \nonumber
\end{eqnarray}
 The variance is computed (as demonstrated, for instance, in Riotto \cite{Riotto}) to be
\begin{eqnarray}
<\zeta^2> = H^2 \frac{(\frac{H}{2 \pi})^2}{(\dot \chi)^2}
\end{eqnarray}

and obtain the power spectrum (using $M_P^{-2} = 8 \pi G_N$) 
\begin{eqnarray}
P_{\zeta} = \frac{1}{2 M_P^2 \epsilon} \bigg(\frac{H}{2 \pi}\bigg)^{2(1-q)} k^{2 q}
\end{eqnarray}
from analyzing the fluctuations of the scalar field, with the Bunch-Davies vacuum, as in Section III.

The rest of the calculation proceeds rather straightforwardly and we get
\begin{eqnarray}
n_{\zeta}-1 = 2q -2 \epsilon (1-q)-2 \epsilon+2 \delta = 2q+2 \eta-6 \epsilon + 2  q \epsilon
\end{eqnarray}

This is a second central result of this paper. There is a correction to the spectral index $n_{\zeta}$, the $2 q$ extra in the sum. Again, in the limit $q \rightarrow 0$, we recover the usual result. This correction would appear for large-$k$, which represent the modes that would have been smaller than the Planck length before and during inflation.

As defined, $q=\frac{4-D}{2}$. It is bigger than $0$ ($0.5 \: to \: 1.5$) for modes that are much shorter in wavelength than the Planck scale. As mentioned earlier, the various holographic models as well as density of state modifications apply for modes with large wave-vector that would have been close to (or at the) the Planck scale during inflation.

\section{Numerical Estimates, Discussion \& Conclusions}
Starting with the Planck length $\sim 10^{-35} m$, if we expect to have $60$ e-folds during inflation, followed by expansion post-inflation by a factor of $10^{30}$, we would see the length expanded by a factor of $10^{56}$. This would lead to fluctuations with a wavelength of $10^{21}$ meters, this corresponds to a length of $10^6$ light-years. This would correspond to $l \sim 10^3-10^4$. This is at the boundary of resolution of the angular distribution seen with Planck's data, but the above results would lead to (on the usual angular distribution plot) an increasing power spectrum ($\sim k^{2q-2(1-q) \epsilon}$). There are other effects, of course, that are complicated to disentangle from this one (Silk damping, etc. \cite{Silk}), so limits could possibly be placed on $q$ as a consequence. Interestingly, the larger error bars at these ranges of $l$ are consistent with {\it rising} CMB spectrum amplitude - more accurate measurements are awaited.

One can ponder the result of the above calculations in a few ways.

It is clear from observation \cite{Wang} that the spectral index for scalar fluctuations is approximately scale invariant, i.e., $n_{\zeta}\approx 0.96$. 

First, the variety of possible values for $q$ and where it starts to become relevant are $q=0.5$ if space were two-dimensional at length scales near the Planck length, to $q=1.5$ in the Banks-Draper ansatz \cite{Banks1}. These are large deviations and would not consistent with observation; therefore, they should be considered excluded -  if we trust the inflation models that need a small $\epsilon$, we would perforce need to exclude such large density of states modifications. Interestingly, if we see a suitably complex ground state for large-$k$, as in the cases studied by M\&B\cite{Brandenberger1}, we could well  allow a larger value of $q$ to compensate for their $n_{\zeta}=0$.

Second, that we can actually have reasonably large values of $q$ (say $q=0.5$), but allow either larger values for $\epsilon$ in a suitably constructed theory of inflation, or a more complex ground state \cite{Brandenberger1}. In the first case, since (with the definition $\epsilon = 3 \frac{ \frac{(\dot \Phi)^2}{2} }{\frac{(\dot \Phi)^2}{2}+V(\Phi)}$), the value $\epsilon=1$ corresponds to the the equation of state parameter $w = \frac{P}{\rho}=-\frac{1}{3}$, so one can allow for values of epsilon almost all the way to $1$ to still maintain the conditions, i.e., $\ddot a>0$, that inflation produces and still solve the flatness and horizon problems. The second possibility is explored in Section III.

There is a third interesting possibility that this allows us to explore the situation with negative $q$ through comparison to experiment. In the simplest theories of inflation, deviations of $q$ away from $0$ appear to be excluded.

A fourth possibility, if one believes that $q$ is indeed non-zero, is that it puts limits on how large the number of e-foldings can be - if $N$ were larger than around 120, then we would definitely see the effect of a non-zero $q$ in the spectral index for simple models of inflation.

In conclusion, we have extended the theory of quantum perturbations due to inflation to account for a varying density of states and have obtained potentially experimentally observable corrections to the running spectral index for short wavelengths. If we trust the simplest models for inflation, the result excludes a reduction of the density of states as well as of dimensionality. It is conceivable that alternative inflation models that can work with larger values of $\epsilon$ would be consistent with a reduced density of states as well as observations.



\section{Acknowledgments}
Discussions with Professor Scott Thomas at the NHETC Rutgers are gratefully acknowledged, as also the hospitality of the Rutgers Physics Department. The perceptive comments of two anonymous reviewers is gratefully acknowledged, especially the pointer to some of the trans-Planckian work.


\begin{thebibliography}{1}
\bibitem{Comment} The Fourier description is $\sigma(\vec k)$ multiplied by a function of $q(\vec k)$. Now $q$ is a function only of the magnitude of $\vec k$; hence the Fourier mode equation can be written as in Equation (5). Further, $ \bigg( (1-q) \frac{a^{''}}{a} - q (1-q)\bigg(\frac{a^{'}}{a} \bigg)^2  \bigg) = \frac{b''}{b}$, where $b=a^{1-q}$, so that the usual calculation can be written in terms of $b$, rather than $a$.
\end{thebibliography}
\end{document}